\documentclass{aastex}
\usepackage{spr-astr-addons}
\usepackage{url}

\RequirePackage{color}

\begin{document}

\title{Compositional waves and variations in the atmospheric abundances
of magnetic stars}
\shorttitle{Compositional waves and variations in the abundances
of stars}
\shortauthors{V.Urpin}

\author{V.Urpin
\altaffilmark{} 
}

\altaffiltext{}{A.F.Ioffe Institute of Physics and Technology,
           194021 St. Petersburg, Russia
}

\begin{abstract}
The stars of the middle main sequence often have relatively 
quiescent outer layers and spot-like chemical structures may 
develop in their atmospheres. Recent observations show that 
abundance peculiarites can change as stars evolve on the main 
sequence and the timescale of these changes lies in a wide 
range from million years to months. These observations imply
that, perhaps, our understanding of diffusion processes at work 
in magnetic stars is incomplete and a more detailed analysis
of these processes is required. In the present paper, we 
consider diffusion caused by a combined influence of the 
electric current and the Hall effect. Such diffusion has a 
number of very particular properties and, generally, can change 
the surface chemistry of stars in combination with other diffusion 
processes. For instant, current-driven diffusion is accompanied 
by a propagation of the special type of waves in which only the 
impurity number density oscillates. Propagation of such waves 
changes the shape and size of spots as well as chemical abundances 
within them. The period of compositional waves depends on the 
parameters of plasma (magnetic field, electric current, 
temperature, etc.) and can be different for the waves of 
different elements. Compositional waves exist in the regions 
where the magnetic pressure is greater than the gas pressure. 
These waves can be the reason of variations with different 
timescales in the abundance peculiarities of magnetic stars.

\end{abstract}


\section{Introduction}

The stars of the middle main sequence have quiescent outer layers, 
and often chemical structures with abundance peculiarities 
may develop in their atmospheres. Likely, diffusion plays an 
important role in the formation of such structures (see, e.g., 
\cite{khoh85}) but many details of diffusion processes are 
still the subject of debate. The point is that diffusion in 
astrophysical bodies is influenced by a number of factors (gravity, 
radiative force, magnetic field, temperature gradient, etc.; see 
\cite{spit98}) and is rather complicated. Therefore, chemically 
peculiar stars are excellent laboratories to study diffusive 
processes in plasma. Note that, apart from the diffusion model, 
some other models have been suggested to account for the observed 
abundance peculiarities in stars (e.g., nuclear evolution, mass 
transfer from a companion, selective accretion from ISM; see 
\cite{land14} for discussion) but all these models are much less 
developed than the diffusion model. 

At present, little is known about the temporal evolution of 
chemical structures during the main sequence evolution of 
stars. Only recently, using high-resolution and high signal-to-noise 
radio spectra, changes in the atmospheric abundances of trace 
elements have been detected in chemically peculiar stars. 
For example, \cite{briq10} discovered noticeable variations 
of $Ti$, $Sr$, and $Y$ with the characteristic timescale $\sim 65$ days
on the surface of HD 11753. Later on, this result was confirmed by 
\cite{korh13} who also detected variations of HD 11753 on 
a long time scale. Note that a slow evolution of chemical spots is 
also seen in AR Aur (\cite{hub10}). A temporal evolution has 
been studied through the main-sequence lifetime of magnetic 
peculiar stars also by \cite{bai14}. These authors have performed 
spectral analysis of 15 Bp stars that are members of open 
clusters and thus have well-known ages. \cite{bai14} considered 
global average abundances in these stars rather than chemical spots 
and discovered a systematic time evolution of trace elements in 
these stars. The authors have attempted to interpret the observed 
variations in the context of radiatively driven diffusion theory but 
not all abundances can be explained in the frame of this theory. 
All these findings indicate the hitherto not well understood physical 
processes that cause a rather fast changes of the chemical spots.

Diffusion of trace elements in plasma can differ qualitatively from 
that in neutral gases because of the presence of electric currents. 
This particularly concerns hydrogen plasma where the rate of 
momentum exchange between electrons and protons is comparable to 
the rate of the momentum redistribution among protons. 
It was first clearly understood by \cite{bra65} in his theory of 
transport phenomena in a high-temperature plasma. This point can 
be clarified by simple qualitative estimates. Indeed, the momentum 
of electrons is $\sim m_e c_e$ ($c_e= \sqrt{k_B T/ m_e}$ is the 
thermal velocity of electrons), and the rate of momentum transfer 
from electrons to protons is $\sim m_e c_e \nu_e \sim m_e c_e / \tau_e$ 
where $\nu_e$ is the frequency of electron collisions with protons
and $\tau_e$ is the electron relaxation time. 
On the other hand, the momentum of protons is $\sim m_p c_p$ where 
$c_p = \sqrt{k_B T/ m_p}$ is the thermal velocity of protons and, 
correspondingly, the rate of momentum redistribution among protons 
is  $\sim m_p c_p \nu_p \sim m_p c_p / \tau_p$ where $\nu_p$ is the 
frequency of proton collisions. Comparing these estimates, we 
obtain that the rate of momentum exchange between electrons and protons 
is of the same order of magnitude: the momentum transfered in a 
collision by electrons is smaller but the frequency of collisions 
is higher. Therefore, the momentum taken from (or transfered to) 
the background plasma by trace elements is redistributed among 
both electrons and protons, and neglecting the electron contribution 
is unjustified in plasma with $Z_0 \sim 1$ where $Z_0$ is the charge 
of the background ions. However, if the background plasma consists 
of ions with $Z_0 \gg 1$, then one should replace $\tau_p$ by the 
relaxation time of the background ions that is $\sim \tau_p / Z_0^4$. 
The rate of momentum transfer by ions turns out to be $\sim Z_0^4$ 
times greater than that by electrons. Therefore, if $Z_0 \gg 1$, 
the electron contribution to momentum transfer is small and can be 
neglected. Note also the crucial importance of the momentum transport 
by electrons in magnetized plasma. Electrons can be magnetized
by an essentially weaker magnetic field than protons or impurities.
Therefore, transport processes are influenced by a relatively weak 
field since electrons give a comparable to 
protons contribution to the momentum transport (see, \cite{urp15}
for detail).

Chemical inhomogeneities can appear in quiescent plasma because of 
a number of reasons (e.g., diffusion in the magnetic field, 
nuclear evolution with non-spherical mixing, accretion from a companion 
or ISM, etc.). The model of diffusion in a star with the magnetic
field seems to be most developed. The magnetic field can magnetize 
electrons and result in anisotropic transport and inhomogeneous distribution 
of trace elements. Anisotropy of diffusion is characterized by the 
Hall parameter, $x_e = \omega_{Be} \tau_e$, where $\omega_{Be} = 
e B/m_e c$ is the gyrofrequency of electrons and $\tau_e $ is their 
relaxation time; $B$ is the magnetic field. In hydrogen plasma, 
$\tau_e = 3 \sqrt{m_e} (k_b T)^{3/2}/4 \sqrt{2 \pi} e^4 n \Lambda $ 
where $n$ and $T$ are the number density of electrons and their 
temperature, respectively, $\Lambda$ is the Coulomb logarithm (see, 
e.g., \cite{spit98}). 

In this paper, we consider one more diffusion process that can 
be responsible for the formation of chemical inhomogeneities and their 
evolution. This process is caused by a combined influence of electric 
currents and the Hall effect and can operate on a timescale shorter 
than the standard diffusion timescale (\cite{urp15}). Using a simple 
model, we show in Sec.2 that the interaction of the electric current 
with impurities leads to their diffusion in the direction perpendicular 
to both the electric current and magnetic field. We argue in Sec.3 that 
such diffusion in combination with the Hall effect can be the reason of 
a particular type of waves in which the impurity number density 
oscillates alone. These waves can be responsible for a change of 
the shape and size of chemical spots and even can lead to migration of 
spots under certain conditions (see discussion in Sec.3 and 4).

\section{Diffusion velocity}

Consider a cylindrical plasma configuration with the magnetic
field parallel to the axis $z$, $\vec{B} = B(s) \vec{e}_{z}$; $(s, 
\varphi, z)$ and $(\vec{e}_s, \vec{e}_{\varphi}, \vec{e}_{z})$ are 
cylindrical coordinates and the corresponding unit vectors. The 
electric current in such configuration is equal to $j_{\varphi} = - 
(c/4 \pi) (d B/d s)$. We use this simplified model to understand 
the main qualitative features of compositional waves. In some cases, 
however, the considered configuration can mimic the real magnetic 
fields. For example, the magnetic field near the magnetic pole is 
close to a cylindrical geometry (see, e.g., \cite{urp93}).
Note that $B(s)$ cannot be arbitrary function of $s$ because
the magnetic configurations can be unstable for some dependences 
$B(s)$ (see, e.g., \cite{tay73}, \cite{bon08a,bon08b}). 

We assume that plasma is fully ionized and consists of electrons 
$e$, protons $p$, and small admixture of heavy ions $i$. The 
number density of species $i$ is small and does not influence  
dynamics of plasma. The partial momentum equations in 
multicomponent plasma have been considered by a number of authors 
(see, e.g., \cite{bra65,vek75,urp81,vek87}). For example, 
\cite{urp81} considers the hydrogen-helium plasma but the derived 
equations describe also the hydrogen plasma with an admixture of 
any heavy ions if their number density is small. If the hydrodynamic 
velocity of plasma is zero and only small diffusive velocities are 
non-vanishing, the partial momentum equation for the species $i$ reads 
\begin{equation}
- \nabla p_i + Z_i e n_i \left( \vec{E} + \frac{\vec{V}_i}{c} 
\times \vec{B} \right) + \vec{R}_{ie} + \vec{R}_{ip} + \vec{F}_i 
= 0, 
\end{equation}
where $Z_i$ is the charge number of ions $i$, $p_i$ and
$n_i$ are their pressure and number density, $\vec{E}$ 
is the electric field, and $\vec{V}_i$ is the diffusion 
velocity. The force $\vec{F}_i$ is the external force on species 
$i$; in stellar conditions, $\vec{F}_i$ is the sum of gravitational 
and radiative forces. Since diffusive velocities are usually very 
small, we neglect the terms proportional to $(\vec{V}_i \cdot \nabla) 
\vec{V}_i$ in the momentum equation (1). The forces $\vec{R}_{ie}$ 
and $\vec{R}_{iH}$ are caused by the interaction of ions $i$ with 
electrons and protons, respectively. Note that forces 
$\vec{R}_{ie}$ and $\vec{R}_{iH}$ are internal and the sum of 
internal forces over all plasma components is zero in accordance 
with Newton's third law. 

For the sake of simplicity, we consider plasma with $T$=const. 
If $n_i \ll n$, then $\vec{R}_{ie}$ is given by      
\begin{equation}
\vec{R}_{ie} = - (Z_i^2 n_i / n) \vec{R}_{e}
\end{equation}
where $\vec{R}_{e}$ is the force acting on the electron gas 
(see, e.g., \cite{urp81}).  
One can use for $\vec{R}_{e}$ the expression for hydrogen 
plasma (\cite{bra65}) since $n_i \ll n$. 

Generally, $\vec{R}_{e}$ contains terms proportional 
to the temperature gradient and relative velocity of $e$
and $p$. If $T=$const, then the expression 
for $\vec{R}_{e}$ reads
\begin{equation}
\vec{R}_{e} = - \alpha_{\parallel} \vec{u}_{\parallel} -   
\alpha_{\perp} \vec{u}_{\perp} + \alpha_{\wedge} \vec{b}
\times \vec{u},
\end{equation}   
where $\vec{u} = - \vec{j}/en$ is the current velocity of 
electrons; $\vec{b} = \vec{B}/B$; the subscripts $\parallel$,
$\perp$, and $\wedge$ denote the parallel, perpendicular, and 
the so called Hall components of the corresponding vector; 
$\alpha_{\parallel}$, $\alpha_{\perp}$ and $\alpha_{\wedge}$ are 
the coefficients calculated by \cite{bra65}. The force 
(3) describes a standard friction force 
caused by a relative motion of electrons and protons in 
magnetized plasma. In our model, $\vec{u} = (c/4 \pi e n) 
(d B /d s) \vec{e}_{\varphi}$ and $\vec{B} \perp \vec{u}$ 
and $\vec{u}_{\parallel} = 0$. In this paper, we consider 
diffusion only in a relatively weak magnetic field with  
$x_e \ll 1$. Substituting $\vec{u}$ into Eq.(3) and using 
coefficients $\alpha_{\perp}$ and $\alpha_{\wedge}$ with 
the accuracy in linear terms in $x_e$, we obtain 
\begin{equation}
R_{ie \varphi} \!\!\! = \! Z_i^2 n_i \left( \! 0.51\frac{m_e}{\tau_e} u 
\! \right), \;\; 
R_{ie s} \!= \!\!\! Z_i^2 n_i \left( 0.21 x \! \frac{m_e}{\tau_e} u 
\! \right). 
\end{equation}

The friction force $\vec{R}_{ip}$ is proportional to the 
relative velocity of ions and protons. Like $\vec{R}_e$, 
this force has a tensor character and depends on $B$.  
However, the dependence of $\vec{R}_{ip}$ on the magnetic 
field is insignificant if $x_e \ll 1$ (see, e.g., \cite{urp81}).
This is qualitatively clear because it is much more dificult 
to magnetize heavy ions than protons or electrons. 
The force $\vec{R}_{ip}$ is particularly simple if $A_i = 
m_i/m_p \gg 1$. Neglecting the influence of 
the magnetic field on 
$\vec{R}_{ip}$ and taking into account that the velocity of 
the background plasma is zero, $\vec{V}_p= 0$, the force 
$\vec{R}_{ip}$ can be written as 
\begin{equation}
\vec{R}_{ip} = (0.42 m_i n_i Z^2_i / \tau_i) (-\vec{V}_i),
\end{equation} 
where $\tau_i = 3\sqrt{m_i} (k_B T)^{3/2} / 4 \sqrt{2 \pi} e^4 n 
\Lambda$; $\tau_{i}/ Z^2_i$ is the timescale of $i-p$  
scattering; we assume that $\Lambda$ is the same for all  
collisions.

After the above simplifications, the cylindrical components 
of Eq.(1) yield
\begin{eqnarray}
- \frac{d}{ds} \!(n_i k_B T) \! + \! Z_i e n_i \left( \! E_s \! + \!
\frac{V_{i \varphi}}{c} B \! \right) \!+\! R_{ie s} \!+\! R_{ip s} 
\!=\! 0, \\
Z_i e n_i \left( \! E_{\varphi} \! - \! \frac{V_{i s}}{c}  B \! \right) + 
R_{ie \varphi} + R_{ip \varphi} = 0, \\
- \frac{d}{dz} (n_i k_B T) + Z_i e n_i E_z + R_{ie z} + R_{ip z} 
+ F_{iz}  = 0.
\end{eqnarray} 
In the chosen magnetic configuration, we have $R_{ie z} = 0$.
Eqs.(6)-(8) depend on cylindrical components of the electric 
field. These components can be determined from the momentum 
equations of electrons and protons
\begin{eqnarray}
- \nabla (n k_B T) - e n \left( \vec{E} + \frac{\vec{u}}{c} \times
\vec{B} \right) + \vec{R}_e + \vec{F}_e = 0, \\
- \nabla (n k_B T) + e n \vec{E} - \vec{R}_e + \vec{F}_p = 0.  
\end{eqnarray}
In these equations, we neglect collisions of electrons and protons 
with the ions $i$ since $n_i \ll n$. The sum of Eqs.(9) and (10) 
yields the equation of hydrostatic equilibrium. The difference of 
Eqs.(10) and (9) yields 
\begin{equation}
\vec{E} = - \frac{1}{2} \frac{\vec{u}}{c} \times \vec{B} +
\frac{\vec{R}_e}{en} - \frac{1}{2en} (\vec{F}_p - \vec{F}_e).
\end{equation}
Taking into account the friction force $\vec{R}_e$  (Eq.(3))
and the coefficients $\alpha_{\perp}$ and $\alpha_{\wedge}$
calculated by \cite{bra65}, we obtain with accuracy in 
linear terms in $x_e$ 
\begin{eqnarray}
E_s \! = \! - \! \frac{uB}{2c} - \frac{1}{e} \! \left( \! 0.21 
\frac{m_e u}{\tau_e} x_e \! \right) , \; 
E_{\varphi} \! =  -  \frac{1}{e} \! \left( \! 0.51 
\frac{m_e u}{\tau_e}  \! \right), \nonumber \\
E_z = - \frac{1}{2en} (F_{pz} - F_{ez}).
\end{eqnarray}
Substituting Eqs.(4) and (12) into Eq.~(8), we obtain the 
following expression for the vertical velocity 
\begin{equation}
V_{iz} =  - D \frac{d \ln n_i}{d z } + \frac{D}{n_i k_B T} 
F_z^{(i)},
\end{equation}
where $D = 2.4 c_i^2 \tau_i / Z_i^2$ is the diffusion
coefficient, $c_i^2 = k_B T/ m_i$, and
\begin{equation} 
F_z^{(i)} = F_{iz} - \frac{Z_i n_i}{2 n} (F_{pz} - F_{ez}). 
\end{equation}
Usually, acceleration due to the radiative energy flux and 
gravitational settling give the main contribution to
$F_z^{(i)}$ (\cite{mich76,vau79,ale06}). The diffusion velocity 
caused by these forces can be relatively fast and, therefore, 
the vertical diffusion often is faster than diffusion parallel 
to the surface. As a result, the vertical distribution of 
trace elements reaches a quasi-steady equilibrium on a relatively short 
timescale. If the radiative and gravitational forces are of 
the same order of magnitude then the velocity of radial diffusion 
can be estimated as $V_r \sim g \tau_i$
(see \cite{vau79}). Then, the chemical equilibrium in the vertical 
direction is reached appoximately on a timescale $\tau_{ver} 
\sim H/V_r \sim H/g \tau_i$ where $H$ is the heigh of the 
atmosphere. Assuming $H \sim 10^8$ cm, $T \sim 10^4$K, $g \sim
10^{4}$ cm/s$^2$, and $n \sim 10^{14}$ cm$^{-3}$, we obtain that 
$\tau_{ver} \sim 10^3-10^4$ yrs if $m_i/m_p \sim 40$. This timescale
is much shorter than the lifetime of peculiar stars. Note that
this timescale can be substantially shorter for stars with 
a hot surface because the radiative acceleration in such stars 
can eccentially excees gravity for some ions.

The tangential components of the difusion velocity can be
obtained from Eqs.~(6)-(7). Taking into acount Eq.~(5)
for $\vec{R}_{ip}$, one can transform Eqs.~(6)-(7) into
\begin{equation}
V_{is} - q V_{i \varphi} = A , \;\;\;\;
V_{i \varphi} + q V_{i s} = G , 
\end{equation}
where
\begin{eqnarray}
A = \frac{D}{n_i k_B T} \left( -\frac{d p_i}{ds} + Z_i e n_i E_s
+ R_{ie s} \right), \\
G = \frac{D}{n_i k_B T} \left( Z_i e n_i E_{\varphi}
+ R_{ie \varphi} \right), \;\;\; q = 2.4 \frac{eB}{Z_i m_i c} \tau_i.
\end{eqnarray}
Then, difusion velocities in the $s$- and $\varphi$-directions 
are 
\begin{equation}
V_{is} = \frac{A + q G}{1+ q^2}, \;\;\;\; 
V_{i \varphi} = \frac{G - qA}{1 + q^2}.
\end{equation}
The parameter $q \sim (m_e/m_i)^{1/2} x_e$ is small even for 
magnetic fields of Ap-stars since $q \ll x_e \ll 1$. Therefore,
$V_{is} \approx A , V_{i \varphi} \approx G$.
Substituting Eqs.(4) and (12) into Eqs.(16)-(17, 
we obtain for the diffusion velocities
\begin{eqnarray}
V_{is}  = V_{n_i} + V_B, \;\;\;
V_{n_i} \!\!=\!\! - D \frac{d \ln n_i}{d s}, \;\;\; 
V_B \! = \! D_B \frac{d \ln B}{d s}; \\
V_{i \varphi} = D_{B \varphi} \frac{d B}{d s}; \;\;\;\;\;\;\;\;\;\;\;\;\;
\;\;\;\;\;\;\;\;\;\;  
\end{eqnarray} 
$V_{ni}$ is the velocity of ordinary diffusion and $V_B$ is the 
velocity caused by the electric current. The 
corresponding diffusion coefficients are
\begin{eqnarray}
D = \frac{2.4 c_i^2 \tau_i}{Z_i^2}, \;\;\; 
D_B = \frac{2.4 c_A^2 \tau_i}{Z_i A_i} (0.21 Z_i - 0.71), \\
D_{B \varphi} = 1.22 \sqrt{\frac{m_e}{m_i}} 
\frac{c (Z_i - 1)}{4 \pi en Z_i}. 
\end{eqnarray}
where $c_i^2 = k_B T/ m_i$ and $c_A^2 = B^2 / (4 \pi n m_p)$. 
Eqs.~(19)-(20) describe the drift of ions $i$ under the combined 
influence of $\nabla n_i$ and $\vec{j}$. 
Note that our consideration of diffusion in plasma differs from that
in astrophysical calculations (see, e.g., \cite{cha70,bur69}) by 
taking into account interaction of protons with electrons.

\section{Compositional waves}

The continuity equation for ions $i$ reads in our model
\begin{equation}
\frac{\partial n_i}{\partial t} + \frac{1}{s} \frac{\partial}{\partial s}
\left( s n_i V_{is} \right) + \frac{1}{s} \frac{\partial}{\partial
\varphi} (n_i V_{i \varphi} )  = 0.
\end{equation}
Consider the behaviour of small disturbances in the impurity 
number density, $n_i$, by making use of a linear analysis 
of Eq.~(23). Assume that plasma is in equilibrium in the 
unperturbed state. Since the number density of trace elements 
is small, their influence on parameters of the basic state is 
negligible. We consider disturbances that do no depend 
on $z$. Denoting the disturbances of $n_i$ by $\delta n_i$ 
and linearizing Eq.(23), we obtain     
\begin{eqnarray}
\frac{\partial \delta n_i}{\partial t} - \frac{1}{s} 
\frac{\partial}{\partial s}
\left( s D \frac{\partial \delta n_i}{\partial s} - s \delta n_i
\frac{D_B}{B} \frac{dB}{ds} \right) +
\nonumber \\
\frac{1}{s} \frac{\partial}{\partial \varphi}
\left( \delta n_i D_{B \varphi} \frac{d B}{d s} \right)= 0.
\end{eqnarray}
We consider disturbances with the wavelength shorter than the 
lengthscale of $B$. In this case, we can use the so called local 
approximation and assume that disturbances
are $\propto \exp(-i k s -M \varphi)$ where $k$ is the wavevector,
$ks \gg 1$, and $M$ is the azimuthal wavenumber. Since the basic 
state does not depend on $t$, $\delta n_i$ can be represented 
as $\delta n_i \propto \exp(i\omega t - i k s -i M \varphi)$ where 
$\omega$ should be calculated from the dispersion equation.
We consider two particular cases of the compositional waves,
$M=0$ and $M \gg ks$.

{\it Cylindrical waves with $M=0$.}
Substituting $\delta n_i$ into Eq.~(24), we obtain
the dispersion equation in the case $M=0$, 
\begin{equation}
i \omega \!=\! - \omega_{R} \! + \! i \omega_{B}, \; \omega_{R} \!=
\!D k^2, \; \omega_{B} \!= \! k D_B (d \ln B / ds).
\end{equation}
This dispersion equation describes cylindrical waves in which 
only the number density of impurity oscillates. The quantity 
$\omega_{R}$ characterizes the decay of waves with the 
characteristic timescale $\sim (D k^2)^{-1}$ typical for a
standard diffusion. The frequency $\omega_{B}$ describes the
oscillation of impurities caused by the combined action of
electric current and the Hall effect. Note that this frequency 
can be of any sign but $\omega_{R}$ is always
positive. The compositional waves are aperiodic if $\omega_{R} >
|\omega_{B}|$ and oscillatory if $|\omega_{B}| > \omega_{R}$. This
condition is equivalent to  
\begin{equation}
c_A^2/c_s^2 > Z_i^{-1} |0.21 Z_i - 0.71|^{-1} kL,
\end{equation}
where $c_s$ is the sound speed, $c_s^2 = k_B T/m_p$.
Compositional waves become oscillatory if 
the field is strong enough and the magnetic pressure is
substantially greater than the gas pressure. The frequency of 
diffusion waves is higher in the region where the magnetic
field has a strong gradient. The order of magnitude
estimate of $\omega_{I}$ yields 
\begin{equation}
\omega_{I} \sim k c_A (1 / Z_i A_i) (c_A / c_i)(l_i / L),
\end{equation}
where $l_i = c_i \tau_i$ is the mean free-path of ions $i$. 
Note that different impurities oscillate with different 
frequences.

{\it Non-axisymmetric waves with $M \gg ks$.} In this case,
the dispersion equation reads
\begin{equation}
i \omega \!= \! - \omega_{R} \!+\! i \omega_{B \varphi}, \;\; 
\omega_{B \varphi} \!= \! (M/s) B D_{B \varphi} (d \ln B / ds).
\end{equation}
Non-axisymmetric waves rotate around the cylindric axis
with the frequency $\omega_{B \varphi}$ and decay slowly on 
the diffusion timescale $\sim \omega_R^{-1}$. The frequency
of such waves is typically higher than that of cylindrical 
waves. One can estimate the ratio of these frequencies as
\begin{equation}
(\omega_{B \varphi} / \omega_B) \sim (B D_{B \varphi} / D_B)
\sim (1 / A_i x_e) (M / ks).
\end{equation}
Since we consider only a weak magnetic field 
($x_e \gg 1$), the period of non-axisymmetric waves
is shorter for waves with $M > A_i x_e (ks)$. The ratio 
of the diffusion timescale and period of non-axisymmetric waves
is 
\begin{equation}
(\omega_{B \varphi} / \omega_R) \sim (1 / x_e) 
(c_A^2 / c_s^2) (Z_i / A_i) (1 / kL)
\end{equation} 
and can be large. Hence, these waves can be oscillatory as well.

\section{Conclusion}

We have considered diffusion of heavy ions under the combined
influence of electric currents and the Hall effect. The velocity 
of such diffusion can be larger than that caused by other diffusion
mechanisms. The considered diffusion forms chemical inhomogeneities 
even if the magnetic field is relatively weak whereas other mechanisms 
require a stronger field (\cite{urp15}). This type of diffusion is 
relevant to the Hall effect and, therefore, it leads to drift of 
heavy ions perpendicular to both the magnetic field and electric current. 

The current-driven diffusion in combination with other diffusion 
processes can contribute essentialy to the surface chemistry of 
various types of stars. Certainly, this type of diffusion may 
play an important role in the surface chemistry of Ap/Bp-stars. 
These stars have a strong magnetic field (see, e.g., \cite{khoh85}) 
that can magnetize the atmospheric plasma and produce a strong Hall 
drift of electrons. These conditions seem to be suitable for the propagation 
of compositional waves considered in this paper and, likely, 
such waves can be the reason of variations in atmospheric abundances 
of these stars. The current-driven diffusion may be also important in 
HgMn stars. The magnetic field of HgMn stars is substantialy weaker 
than that of Ap/Bp stars (see, e.g., \cite{wade04,hub01,hub06}) and, 
likely, it cannot magnetize heavy ions and form element spots by the
mechanism based on the magnetization of such ions. The current-driven 
diffusion requires, however, a relatively weak magnetic field compared 
to other diffusion mechanisms (see discussion in \cite{urp15}). 
Perhaps, the considered diffusion mechanism can be impotant also 
in compact stars like white dwarfs or neutron stars. For instance, 
many white dwarfs and neutron stars have a strong magnetic field 
with complex topology and spot-like structures at the surface. Such 
magnetic configurations can lead to the formation of a spot-like element 
distribution at the surface. Particularly, this 
concerns the accretion phase in binary systems. Many white dwarfs and 
neutron stars in relatively close binaries pass through the accretion 
phase when plasma of a companion is accreted by the compact star. 
Such accretion phase is typical for binaries with various types of  
the companion (see, e.g., \cite{urp98a,urp98b}) and the evolution of 
compact stars is very complicated during this period. For example, a 
strong magnetic field can channel plasma of a companion onto the 
surface of the neutron star and form chemical spots there. The evolution of
such spots on neutron stars can be very complicated because of 
strong magnetic field, high gravity and large luminosity. The 
chemical processes in these spots may influence the burst 
activity of neutron stars (see, e.g., \cite{bro02,cha04}), their 
thermal evolution, etc. and, likely, the current-driven diffusion can 
play an important role in the evolution of spots.

Our study reveals that the particular type of waves (compositional 
waves) may exist in multicomponent plasma in the presence of electric 
currents. Such waves exist only if the magnetic pressure is greater 
than the gas pressure. The compositional waves are slowly decaying 
and characterized by oscillations of the impurity number density alone. 
For example, in subsequent spectral observations, the propagation 
of these waves can manifest themselves by changes in the location of 
regions with a higher intensity in spectral lines (bumps). These bumps 
indicate chemical spots on the surface of peculiar stars and their motion 
can be caused by compositional waves. Note that, generally, bumps in 
different spectral lines can move with different velocities.

The frequency of compositional waves turns out to be 
different for different sorts of ions. Therefore, the abundances 
of different elements in peculiar stars can vary on different 
timescales. The period of waves can change in a wide range from 
that comparable to the lifetime of a star to a rather short value
of the order of few months. 

For instance, the compositional modes can be responsible for
changes in the distribution of few elements on the surface of the 
HgMn star HD 11753 discovered by Briquet et al. (2010). Spectral line 
profile changes were detected by making use of the Doppler imaging 
technique and using two datasets separated by $\sim 65$ days. The 
results of this study revealed noticeable changes in the distribution 
of TiII, SrII, and YII indicating a rather fast chemical spots 
evolution. If this evolution is caused by a propagation of the 
compositional waves then one can expect spectral line profile changes 
on the timescale $\sim \omega_B^{-1}$ (see Eq.(25)). The frequency 
$\omega_B$ can be estimated as $\omega_B \sim (k / L) D_B$ where $L$ 
is the magnetic lengthscale. Substituting the value of $D_B$,
we obtain
\begin{equation}
\omega_B \sim \frac{2 \times 10^{-13}}{\sqrt{A_i}} 
\frac{B_{100}^{2}}{L_{10} \lambda_{10} \rho_{-10}^2} 
\frac{T_4^{3/2}}{\Lambda_{10}} \;\; {\rm s}^{-1}, 
\end{equation}
where $B_{100} = B/100$ G, $L_{10} = L/10^{10}$ cm, $lambda_{10} =
(2 \pi/k)/10^{10}$ cm, $\rho_{-10} = \rho / 10^{-10}$ g/cm$^3$,
$T_{4} = T/10^4$ K, and $\Lambda_{10} = \Lambda/10$. Let us assume
in estimates that $T_4 =1.5$, $\Lambda=3$, and the wavelength of
compositional waves, $\lambda$, is comparable to the magnetic 
lengthscale $L$, then the characteristic timescale of chemical 
variations is
\begin{equation}
\tau \sim \frac{1}{\omega_B} \sim 10^{13} \sqrt{A_i} L_{10}^2
\frac{\rho_{-10}^2}{B_{100}^2} \;\;\;s.
\end{equation}
According to Briquet et al. (2010) the log of the abundance of
hydrogen is equal to 12.0 in the region of a fast chemical evolution.
Although strong magnetic fields have not generally been found in
HgMn stars, it has never been ruled out that these stars might have
tangled magnetic fields $\sim$(a few thousand Gauss) with no net
longitudinal component (see, e.g., \cite{hub01}). Therefore, one
cannot rule out that HD 11753 has a transverse magnetic field of
such order of magnitude. We assume in estimates that $B=3 \times 
10^3$ G. If $L_{10} =1$ then we obtain
\begin{equation}
\tau \sim 10^6 \sqrt{A_i} \;\;{\rm s}  \sim 10 \sqrt{A_i} \;\;
{\rm days}.
\end{equation}   
This estimate is in a good agreement with the characteristic timescale
measured by Briquet et al. (2010). Note that our model predicts that
the timescale $\tau$ can be different for different elements and $\propto
\sqrt{A_i}$. The square root of atomic numbers for TiII:SrII:YII is equal 
approximately to 6.9:9.4:9.4. Therefore, the timescales for SrII and YII
should be approximately the same according to our model, whereas chandes
of TiII can occur on a timescale about 20-30\% shorter. Since Briquet et
al. (2010) compared only two datasets separated by 65 days, more
detailed study of the abundance variations in HD 11753 is required.

\acknowledgments
The author thanks the reviewer, Dr Jeffrey Bailey, for useful comments
and suggestions. This work was supported by Russian Academy of Sciences 
(Programme OFN-15).

\nocite{*}
\bibliographystyle{spr-mp-nameyear-cnd}
\bibliography{biblio-u1}

\end{document}